\documentclass{osa-article}

\journal{ao}
\articletype{Research Article}
\usepackage{lineno}
\usepackage{wrapfig}

\begin{document}

\title{Imaging Scatterometer for Observing In-Situ Changes to Optical Coatings During Air Annealing}

\author{Michael Rezac$^1$, Daniel Martinez$^1$, Amy Gleckl$^1$, Joshua R. Smith$^{1,*}$}

\address{$^1$ The Nicholas and Lee Begovich Center for Gravitational-Wave Physics and Astronomy, California State University, Fullerton, US}
\email{$^*$josmith@fullerton.edu}



\begin{abstract}
Annealing of amorphous optical coatings has been shown to generally reduce optical absorption, optical scattering, and mechanical loss, with higher temperature annealing giving better results. The achievable maximum temperatures are limited to the levels at which coating damage, such as crystallization, cracking, or bubbling will occur.
Coating damage caused by heating is typically only observed statically, after annealing. 
An experimental method to dynamically observe how and over what temperature range such damage occurs during annealing is desirable as its results could inform manufacturing and annealing processes to ultimately achieve better coating performance.  
We developed a new instrument that features an industrial annealing oven with holes cut into its sides for viewports to illuminate optical samples and observe their coating scatter and eventual damage mechanisms in-situ and in real-time during annealing. We present results that demonstrate in-situ observation of changes to titania-doped tantala coatings on fused silica substrates.  We obtain a spatial image (mapping) of the evolution of these changes during annealing, an advantage over x-ray diffraction, electron beam, or Raman methods. We infer, based on other experiments in the literature, these changes to be due to crystallization. We further discuss the utility of this apparatus for observing other forms of coating damage such as cracking and blisters. 
\end{abstract}
\section{Introduction}
\label{sec:intro}

Optical interference coatings with as low as possible optical and mechanical losses are in demand for high-precision optical measurement applications such as atomic clocks and interferometric gravitational-wave detectors~\cite{Harry:2011book}.
In gravitational-wave observatories such as Advanced LIGO~\cite{TheLIGOScientific:2014jea} and Advanced Virgo~\cite{Acernese_2014}, optical and mechanical losses of the coatings must be very low in order to not degrade the detector sensitivity. 
Scatter from gravitational-wave detector optics increases the quantum-noise limit to sensitivity, leads to stray light that causes nonlinear noise by coupling back into the main beam after scattering off moving elements~\cite{Flanagan_1994, Accadia_2010}, and degrades squeezed states of light~\cite{Kwee:2014vba}.  
Optical absorption drives thermal effects, such as thermal lensing, within the gravitational-wave detector optics which can degrade detector sensitivity and performance~\cite{Wang_2017}.
Mechanical loss determines the off-resonant Brownian motion of optical coatings (also known as coating thermal noise)~\cite{Harry:2011book}, which is a limiting noise source in gravitational-wave detectors~\cite{PhysRevD.102.062003}. 
Achieving the low optical and mechanical loss of coatings for gravitational-wave detectors is accomplished by selecting materials with excellent optical and mechanical properties, using ion-beam sputtering deposition with closely controlled deposition temperature and energy, ensuring cleanliness and purity, and through post-deposition annealing. 

Currently, Advanced LIGO and Advanced Virgo use coatings formed by TiO$_2$-doped Ta$_2$O$_5$ (high index) and SiO$_2$ (low index) ion-beam sputtered layers produced at Laboratoire des Matériaux Avancés. 
Post-deposition annealing of such coatings to 600$^{\circ}$C, in air, has been shown to reduce their scatter~\cite{Sayah:21,Capote:21}, absorption~\cite{Fazio:20}, and mechanical loss~\cite{Granata:2019fye}, with higher temperatures (in general) giving better results.
The titania dopant is added to further decrease the mechanical loss and the absorption of the Ta$_2$O$_5$ layers~\cite{Granata:2019fye}.
We note that other dopants such as zirconia can be added to frustrate crystallization in tantala~\cite{Abernathy_2021} and titania-doped-tantala~\cite{doi:10.1116/6.0001074}, though the samples used here do not include zirconia. 
There is currently heavy research into identifying coatings for future detectors that will have even lower coating thermal noise~\cite{2018RSPTA.37670282S} and as good optical properties as the current coatings. Post-deposition annealing of materials such as TiO$_2$-doped GeO$_2$ with SiO$_2$ is one path that has already shown promise~\cite{PhysRevLett.127.071101}. 

The practical limits to the maximum annealing temperatures achievable for ion-beam-sputtered amorphous thin-film coatings are determined by the onset of crystallization or damage mechanisms such as delamination, blisters, and cracks. 
The presence of such damage is often observed using optical methods such as visual inspection, imaging with a scatterometer or a microscope, and x-ray diffraction for crystallization. 
Typically, inspection for damage associated with annealing is performed before and after a given annealing regimen, providing incomplete information about the conditions that lead to the onset and growth of damage. 
Previous work by our group introduced an in-situ method to observe optical scattering from coatings while they are annealed in vacuum and showed that scattered light from  TiO$_2$:Ta$_2$O$_5$, decreases during annealing to 500$^{\circ}$C in vacuum~\cite{Capote:21}.  
Thus far, achieving much higher temperatures with this setup has proven challenging. 
Furthermore, air annealing is more commonly used in the gravitational-wave optics community than vacuum annealing and has the advantages of demonstrated improvement of mechanical and absorption losses. 
Here, in Section~\ref{sec:setup}, we describe a new instrument that was developed to meet the goal of imaging scattered light from the coatings while they are being annealed in air to temperatures of 800$^{\circ}$C, or higher. 
In Section~\ref{sec:results} we show that this instrument is capable of imaging the onset and growth of crystals in samples with single-layer coatings of TiO$_2$:Ta$_2$O$_5$ (described in Section~\ref{sec:samples}) and the onset and growth of blisters in TiO$_2$:GeO$_2$/ SiO$_2$. 
In Section~\ref{sec:conclusion}, we discuss ways that this instrument will provide deeper insight into crystallization and other coating damage mechanisms by measuring their onset and evolution versus temperature.

\section{Experimental Setup and Procedure}
\label{sec:setup}

\begin{figure}[ht]
  \centering
  \includegraphics[width=\linewidth]{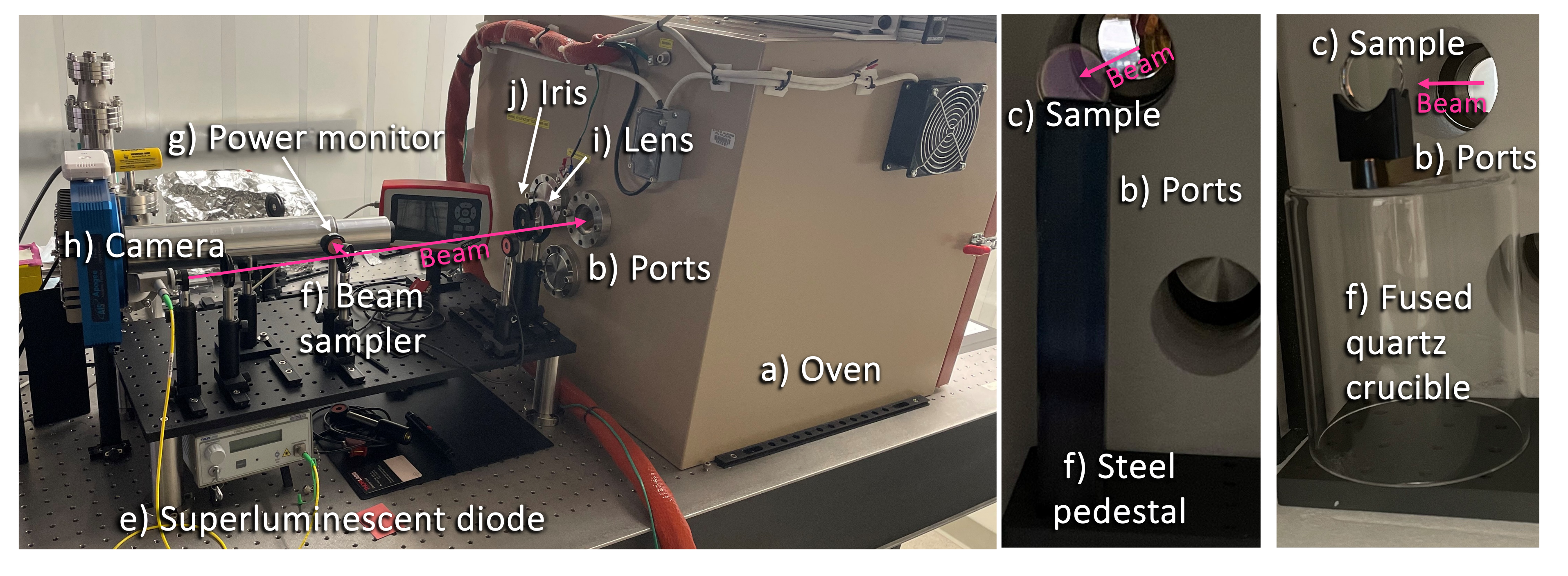}
  \caption{Setup of the Air Annealing Scatterometer. Labeled components are described in the text. \textit{Left:} setup exterior to the oven. \textit{Middle:} setup inside the oven used for the first four samples. \textit{Right:} setup inside the oven for the last sample, where the stainless steel pedestal was replaced by a fused quartz crucible to decrease thermal expansion. The thermocouple is located just to the right of the viewport shown, on the ceramic interior surface. 
  \label{fig:setup}}
\end{figure}

The setup of the Air Annealing Scatterometer (AAS) is shown in Figure~\ref{fig:setup}. The basic operation is as follows. A sample coated optic is mounted within the oven, then illuminated by an external light source and imaged at regular intervals (once per minute here) using a CCD camera while an annealing temperature profile is carried out. 
This system was designed to meet the stated goal of imaging coating scatter in-situ to 800$^{\circ}$C, or higher. The components that were required to meet that goal and their interplay are described below.

The setup requires  a programmable oven (\textbf{a}) capable of reaching 800$^{\circ}$C and of accurately carrying out heating profiles with multiple ``ramp" (increase/decrease temperature) and ``dwell" (maintain constant temperature) segments.
We chose the industrial annealing oven ST-1500C-121012 by SentroTech, which uses molybdenum disilicide (MoSi$_2$) heating elements and thick ceramic insulation to reach, for an unmodified oven, 1500$^{\circ}$C. 
We worked with Sentrotech's engineering team to add observation and instrument ports (\textbf{b}) to both the front (the side with the dark orange door) and rear of the oven. 
These ports use conflat (CF) flanges to allow the use of heat tolerant gaskets and the option to connect various commercially available components, such as viewports and flanges. 
Because of these holes through the outer walls and insulation, this modified oven's maximum temperature will be reduced to 900$^{\circ}$C-1100$^{\circ}$C. 
We also added an air circulation fan option to the interior of the oven to ensure temperature consistency. 
The temperature is read by an S-type thermocouple that communicates with the oven's controller. 
The thermocouple is located between the two upper ports sticking out from the ceramic into the interior of the oven (nearby and to the right of sample holder shown in Figure~\ref{fig:setup}). 
The oven's controller (Nanodac from Eurotherm) uses proportional–integral–derivative (PID) control and provides an interface for creating heating profiles with up to 30 (ramp, dwell, or target temperature) segments using the software package iTools. 

A coated sample optic (\textbf{c}) is mounted within the oven on a solid stainless steel pedestal (\textbf{d}) (later replaced by a fused quartz crucible) so that it can be both illuminated and imaged through a viewport on the back of the oven.
To most closely match the use case of coated optics in interferometric gravitational-wave detectors, the setup requires monitoring scattered light from samples illuminated at normal incidence by a light source similar in wavelength to that used by LIGO and Virgo (1064\,nm). 
To avoid time-dependent speckle effects associated with coherent light~\cite{Bhandari:11,Kontos:21,Capote:21}, thus better allowing association of small changes in scatter with physical changes in the coatings, we use a 1050\,nm superluminescent diode (SLD) (\textbf{e}, Thorlabs S5FC1050P, with 50\,nm bandwidth and coherence length $L_c=\lambda^2/\Delta\lambda\approx 20$\,$\mu{m}$. 
To monitor fluctuations in the incident power, a few percent of the SLD's output is picked off by a beam sampler (\textbf{f}, Thorlabs BSF10-C) and recorded by a calibrated power meter (\textbf{g}, Thorlabs PM100D). The transmitted light is (optionally) measured by a second power meter after passing the viewport on the front door of the oven. The setup is thus capable of measuring in-situ transmittivity and could be modified to record in-situ reflectivity if desired. 

A low-noise and high resolution camera is required to image the light scattered from the coated optic and identify defects and damage mechanisms such as point scatterers, blisters, and crystals. 
The AAS uses a cooled 4096x4096-pixel astronomical CCD camera (\textbf{h}, Apogee Alta F16M) with programmable capture, adjustable exposure times, and high linearity over a large illumination range. An image of the sample's coated surface at a scattering angle (defined as the angle between the sample's normal and the measured scattered light) of $\theta_s=8^{\circ}$ 
is cast on the CCD chip, with 2X magnification (M=$2.02\pm.03$), using  a single ($f$=200\,mm) converging lens (\textbf{i}) and an adjustable iris (\textbf{j}). 
As the CCD sensor size is a 3.68cm x 3.68cm square, this magnification gives a field of view (at the object plane) of height $1.825+/-.025$\,cm and width $\cos{\theta_s}\approx 0.99$ times that. 
The SLD beams, camera, and imaging optics are all at the same height (i.e., in the plane of the SLD beam).
A narrow-band filter (Edmunds 1050nm/50nm) is installed at the front of the lens tube to limit thermal radiation from the oven's heaters and room light from entering the camera, while allowing the SLD wavelengths to pass. 
To further limit the effects of thermal radiation and to account for ``hot" pixels in the camera, for each ``bright" image that is taken with the SLD illumination on, a ``dark" image is also taken with the SLD off, that can be subtracted from the bright image during analysis. 
Images are recorded using the Flexible Image Transport System (FITS) format, which incorporates metadata such as time stamp, exposure time, camera temperature, and saturation levels.  

A LabView Virtual Instrument (VI) is used to automatically control and acquire data from the SLD, oven, camera, and power monitors. 
A typical experiment goes as follows. A sample is installed and the focus of the imaging optics is checked. 
The desired heating profile, with typical duration of 1-2 days, is created using iTools and loaded to the oven's controller. 
The VI is configured with the desired camera exposure time and imaging cadence. 
The VI is started and it executes the following sequence (where times indicate the duration spent in each state): 
i) read oven set point, heater power, and thermocouple temperature ($<$1\,s);
ii) turn SLD on (1.5\,s); 
iii) read incident and transmitted power monitors ($<$1s); 
iv) bright image exposure (5\,s); 
v) transfer bright image (20\,s); 
vi) turn SLD off (1.5\,s); 
vii) dark image exposure (5\,s); 
viii) transfer dark image (20\,s); 
ix) wait (roughly 9\,s) until the total elapsed time of the entire sequence reaches 60\,s, then repeat. 
In this way, one bright image and dark image along with one data point each for incident and transmitted laser power and oven temperature is collected per minute. 
The images and data are written to disk on the PC running LabView and backed up for analysis.
The setup is located on a passive seismic isolation optical bench, within a laminar flow softwall cleanroom and the room is kept closed and dark during measurement. A PC in an adjacent room can be used to view the images in real time as they are collected. 

Scattered light is commonly quantified using the Bidirectional Reflectance Distribution Function (BRDF)~\cite{Stover:2012book}. 
\begin{equation}
    BRDF = \frac{dP_s/d\Omega_s }{P_i \cos \theta_s} \cong \frac{P_s/\Omega_s }{P_i \cos \theta_s},
\end{equation}
where P$_i$ is the incident laser power and P$_s$ is the scattered light power measured at polar angle $\theta_s$ by the imaging system, which subtends a solid angle $\Omega_s$.

Analysis of the AAS data to obtain BRDF is accomplished using a custom-written Matlab script which proceeds as follows for each data point (i.e., each set of  bright image, dark image, temperature and power readings for a given time).
While BRDF in general represents angle-resolved scatter, the measurements presented here yield a BRDF only at the fixed angle $\theta_s=8^{\circ}$. 
An elliptically shaped region of interest is defined, by inspecting the bright image, to enclose the coating area that has the beam spot and thus significant scattered light.
The dark image is subtracted from its corresponding bright image. 
The counts of all pixels within the region of interest in the subtracted region are summed up and normalized by the exposure time and the incident power and multiplied by a calibration factor (previously determined by comparing the scattered light from a diffuse reference sample measured by a calibrated power meter and by the CCD camera) to give BRDF~\cite{Magana-Sandoval:12}.  
To better estimate the optical scatter enclosed by the region of interest, 5 larger concentric regions of interest are defined around the first. Their counts are also summed and normalized and then all six values are fit with a line versus pixel area. The resulting y-intercept is taken as the true enclosed BRDF without any additional diffuse light~\cite{Capote:21}. This procedure is especially important to avoid overestimating BRDF when the scatter (at the beam spot) that is enclosed by the inner region is not much brighter than the background in the images. 

While commissioning the AAS instrument, we identified a second illumination and measurement channel that provides complimentary information.
At elevated temperatures, thermal radiation from the heating elements provides an alternative side illumination of the coatings that is particularly useful for viewing blisters and delamination, especially in the ``dark" images with the SLD off. 
While this was not the primary aim of this instrument one result demonstrating this capability is presented in the next section. 
We note that since thermal radiation is so strongly temperature dependent, a more constant dedicated side illumination from an SLD could be added to achieve good damage visibility for the full duration of the experiments.

\section{Samples}
\label{sec:samples}

\begin{figure}
  \centering
  \includegraphics[width=\linewidth]{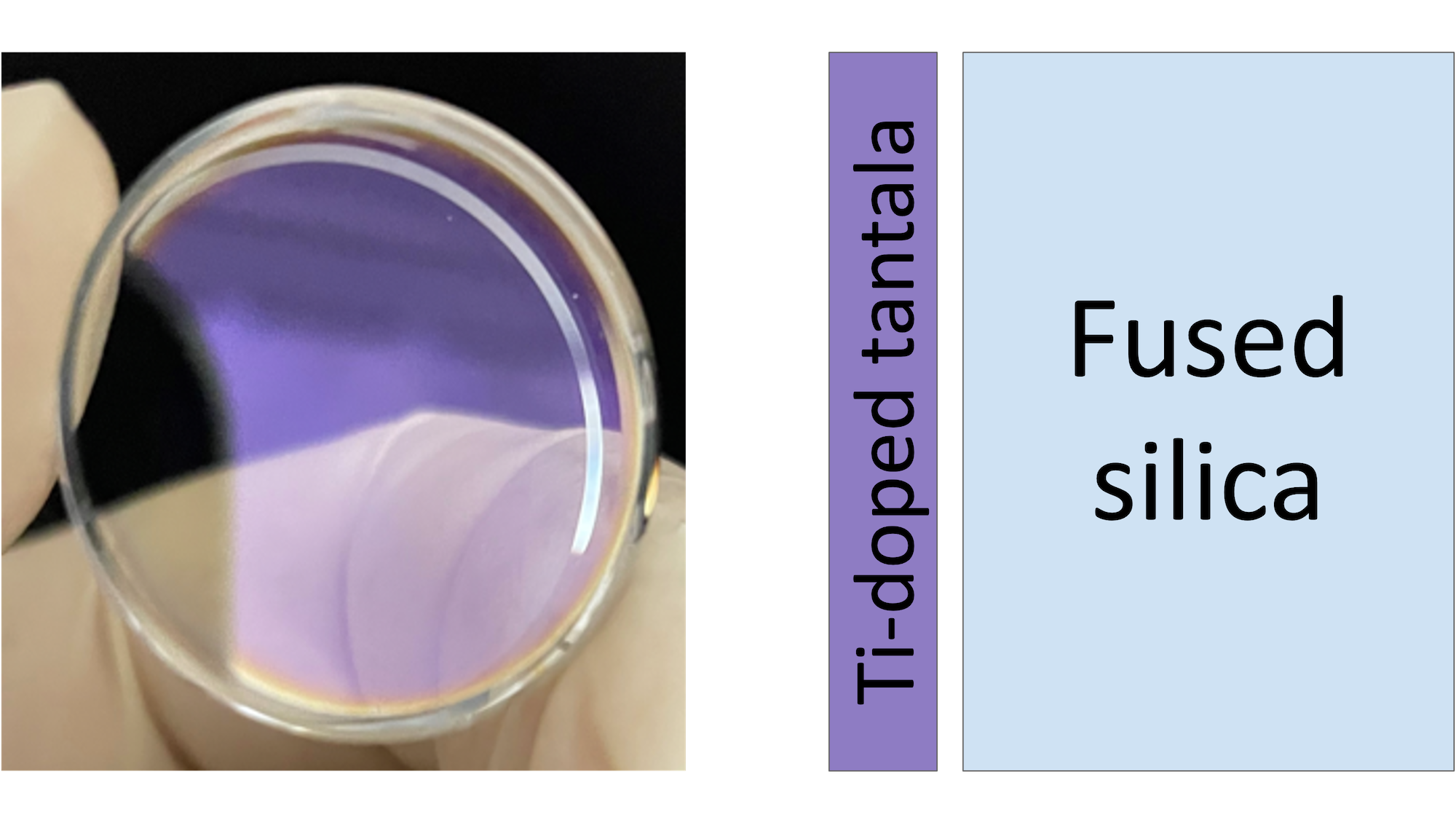}
  \caption{Left: Sample PL003, post-annealing, shown as an example. A single quarter-wavelength-thick (for 1064\,nm, 126\,$\mu$m physical thickness) layer of TiO$_2$:Ta$_2$O$_5$ (Ti/Ta=0.27) is coated (by Laboratoire des Matériaux Avancés) on a superpolished ($\sigma < 0.1$ nm) Corning 7979 fused silica substrate (from Coastline Optics). Right: Diagram (not to scale) of the single-layer coating and substrate (standard 1-inch optic with 25.4\,mm diameter and 6.35\,mm thickness). Five nominally identical samples were used in this study. \label{fig:samples}}
\end{figure}

Five nominally identical samples, see Figure~\ref{fig:samples}, were used in this study.  
The substrates were Corning 7979 fused silica, superpolished by Coastline Optics to $\sigma < 0.1$ nm RMS surface roughness (as measured by Coastline's Zygo 5500 optical profiler), with a 10-5 scratch-dig in the central 80\% of their face surface. The optic barrels were standard polished in order to stop them strongly scattering stray light and thus glowing brightly in the images. The edges were chamfered to avoid accidental chipping.
Coastline produced 20 such samples with serial numbers 7979FSPL001 through 7979FSPL020. Samples (dropping the prefix) PL001, PL003, PL005, PL006, and PL007 were used in this study.
All samples were coated in a single run by Laboratoire des Matériaux Avancés with a single ion-beam-sputtered layer of TiO$_2$:Ta$_2$O$_5$ (dopant level Ti/Ta=0.27) with a quarter-wavelength thickness (126\,$\mu$m physical thickness, assuming n=2.11) for 1064\,nm light. 
Thus these layers are produced by the same vendor and use the same titania-doped-tantala material as is used for the high-index of refraction layers in the current LIGO and Virgo optical coatings~\cite{Granata:2019fye}. 

\section{Results}
\label{sec:results}

\begin{figure}
  \centering
  \includegraphics[width=0.8\linewidth]{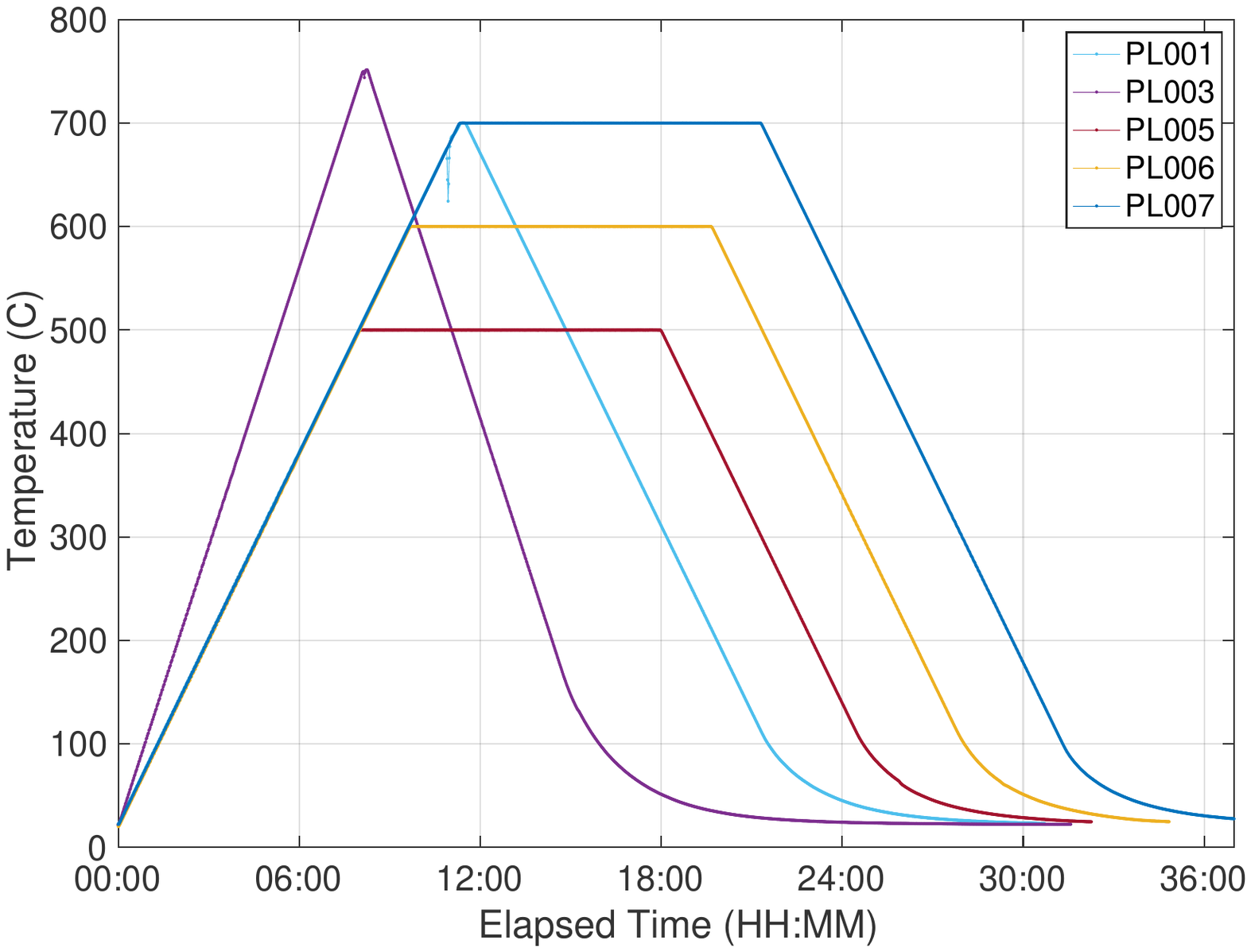}
   \includegraphics[width=0.8\linewidth]{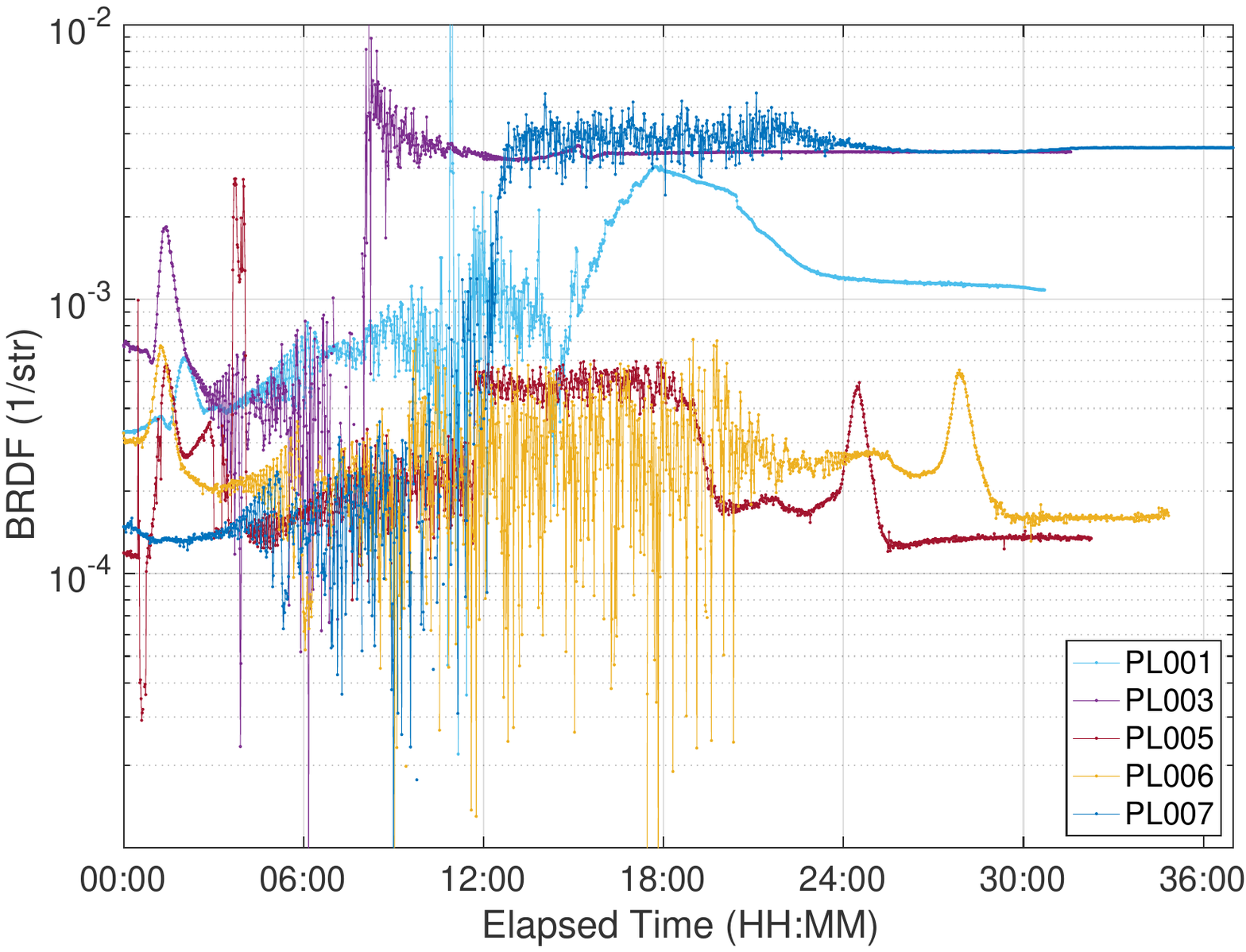}
  \caption{Top: The temperature profiles, with 1C/min ramp rate and variable soak duration, for each sample shown over 36 hours. Cooling rates below 100C do not follow the desired rate because there is no fast cooling fan so cooling relies entirely on radiation. Small spikes in the temperature are due to automation or communication errors. Bottom: BRDF for each sample over the same timescale. Samples PL001 (700C), PL003 (750C), and PL007 (700C) show crystallization. Samples PL006 (600C) and PL005 (500C) show an overall decrease in scatter. The large variations seen at high temperatures are believed to be caused by fluctuations in thermal radiation of the heaters and we are working to eliminate this issue. \label{fig:all-results}}
\end{figure}

The top panel of Figure~\ref{fig:all-results} shows the PID-controlled temperature, measured by the S-type thermocouple mounted inside the oven, versus elapsed time for all five samples. The bottom panel shows the measured BRDF of the coating surface scatter of each sample at $\theta_s=8^{\circ}$ versus the same elapsed time. 
Sample PL003 was ramped to 750$^{\circ}$C with a 1.5$^{\circ}$C/minute rate and a 10-minute soak. This sample crystallized and is described in more detail below. All other samples were ramped at 1.0$^{\circ}$C/minute rate. 
Sample PL001 was ramped to 700$^{\circ}$C with a 10-minute soak and experienced only partial crystallization. 
Samples PL005 and PL006 were ramped to 500$^{\circ}$C and 600$^{\circ}$C, respectively, each with a 10-hour soak. Neither of the samples showed any signs of crystallization in the BRDF or the images. Instead, they both exhibited a decrease in scatter from the start to end of annealing, in agreement with previous work~\cite{Capote:21, Sayah:21}. 
Sample PL007 was ramped to 700$^{\circ}$C with a 10-hour soak and strongly crystallized, as described below. 

Runs with PL001, PL003, PL005, and PL006 used a stainless steel pedestal to support the optic holder at the height of the viewport. The thermal expansion of this pedestal caused the optic to translate up and down by more than 1\,mm during those runs. 
Since the beam and the imaging optics were fixed, this motion caused any bright point scatterers to translate through the beam intensity profile, causing shifts in the BRDF seen as the large bumps during the ramp up and ramp down for those traces. 
For the PL007 run, a fused silica crucible (Advaluetech FQ-2500) was used as the pedestal, greatly reducing the thermal expansion and thus essentially eliminating any translation of the sample during heating. For this sample, the copper gasket for the CF-flange used for the incident light and imaging was removed and this also reduced some stray light artifacts.    

The BRDF for all samples shows a large ``noisy" variation at higher temperatures, above 400$^{\circ}$C, with some data points not shown on the logarithmic scale as they were negative (meaning that the region of interest in the ``bright" image encloses less light than in the ``dark" image). 
This issue was associated with ``flashing" observed in the images 
on the timescale of tens of seconds which causes successive bright and dark images to have differing amounts of background light.
The flashing is due to varying thermal radiation from the heater elements driven by the facts that the thermal irradiance varies as $T^4$ and the heating elements experience much higher temperature variations than the air and other components in the oven. 
Following the measurements presented here, several experiments have been conducted to learn more about this issue. 
The heaters were observed with a secondary camera with higher frame rate video, confirming the flashing and changes were made to the PID parameters to attempt to change the rate of heater switching. 
The issue has not yet been solved. 
As a workaround we have machined stainless steel radiation blocks to surround the optic that should passively lowpass the thermal radiation. 

\subsection{TiO$_2$:Ta$_2$O$_5$-coated sample PL003}

\begin{figure}
  \centering
  \includegraphics[width=1.0\linewidth]{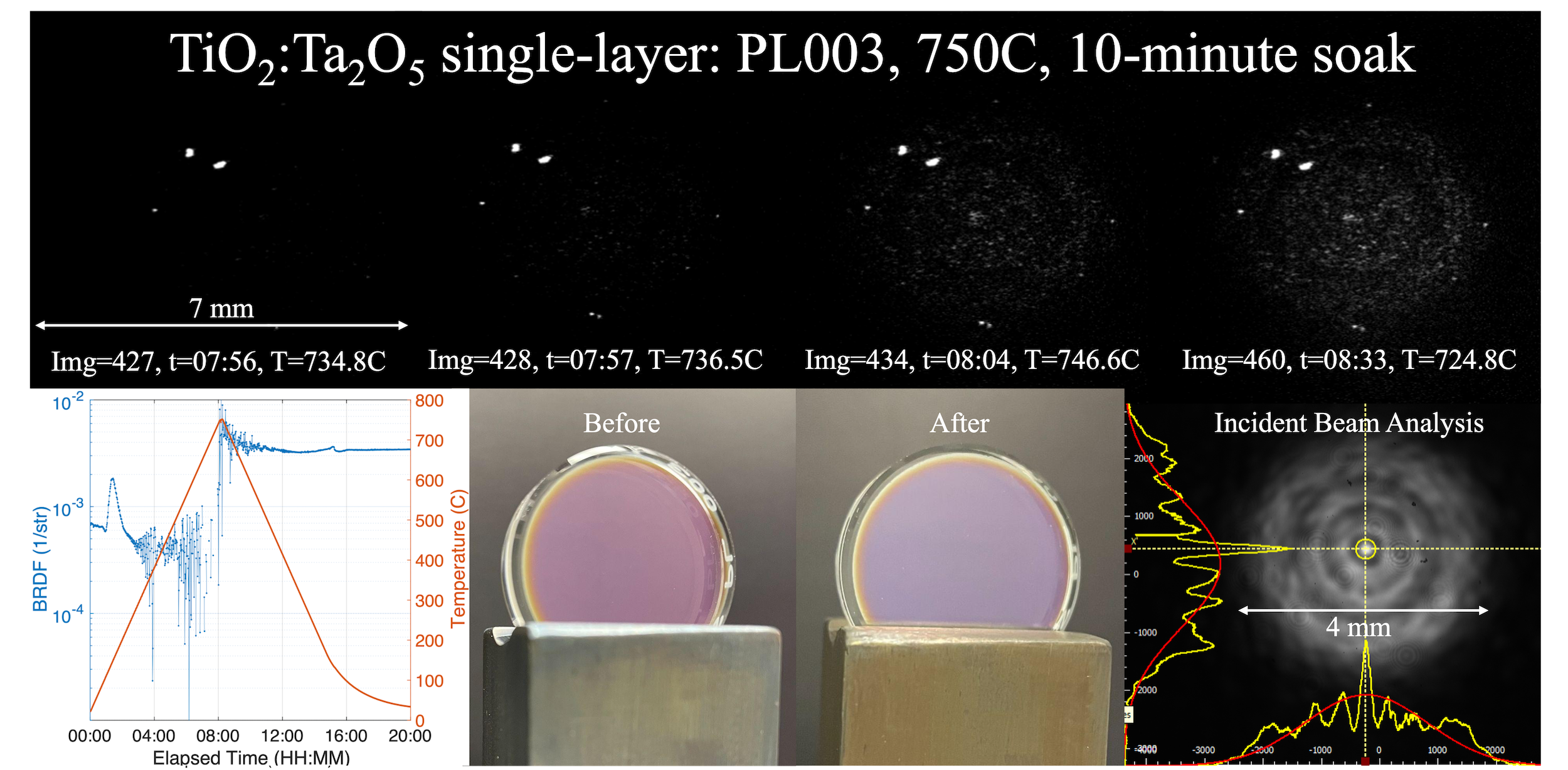}
  \caption{Results of annealing sample PL003, a single quarter-wavelength layer of TiO$_2$:Ta$_2$O$_5$ on a superpolished fused silica substrate. \textit{Bottom, left to right:} Measured oven temperature (orange) and BRDF (blue) of the sample, both versus elapsed time; sample before annealing; sample after annealing, showing the coating is slightly milky and the stainless steel holder changed color; intensity profile of the incident laser beam at the location of the sample (measured with a Thorlabs BC106-VIS beam profiler inserted close to the entry viewport prior to measurements), showing a beam diameter 4\,mm, and for this run, a strongly diffracted beam. The top row shows cropped images of a (7\,mm-wide) region of the sample illuminated by the SLD. \textit{Top, left to right:} At 734.8$^{\circ}$C some point scatterers are visible, but no crystallization is seen; at 736.5$^{\circ}$C weak diffuse scattering from the onset of crystallization is seen; at 746.6$^{\circ}$C the beam intensity profile is scattered quite uniformly by the crystallized coating; at 724.8$^{\circ}$C on the ramp down, the same pattern is seen more brightly.  \label{fig:pl003-results}}
\end{figure}

Figure~\ref{fig:pl003-results} shows the results of annealing sample optic PL003. This simple heating profile was chosen to achieve crystallization in the coating layer based on previous studies~\cite{Fazio:20}. 
The bottom left panel shows the time evolution of the oven's temperature and the optic's bidirectional reflectance distribution function (BRDF). 
The BRDF exhibits a first bright peak at low temperature due to point scatterers getting very bright as the sample moves up (due to thermal expansion of the tall stainless steel holder) and the beam sweeps over them.
At mid-temperatures the aforementioned noisy measurements of BRDF are evident, due to both variations in the heater element thermal radiation and the sample translating within the beam pattern.
Starting around 735$^{\circ}$C there is a factor of ten, and permanent, increase in BRDF which has both the onset temperature and ``frosted" appearance expected for crystallization (and verified elsewhere with, e.g., x-ray diffraction~\cite{Fazio:20}).
Thus, we infer this behavior to be related to crystallization. 
The images along the top row show increasing scattering with time as the crystals grow, with the final image resembling the incident beam pattern shown in the bottom right. For this run, the beam pattern exhibited a strong circular diffraction pattern that was caused by a small iris (to reduce this, the iris was opened more widely for other samples). The before and after visible light images show that the coating has become slightly cloudy.

\subsection{TiO$_2$:Ta$_2$O$_5$-coated sample PL007}

\begin{figure}
  \centering
  \includegraphics[width=1.0\linewidth, angle=0]{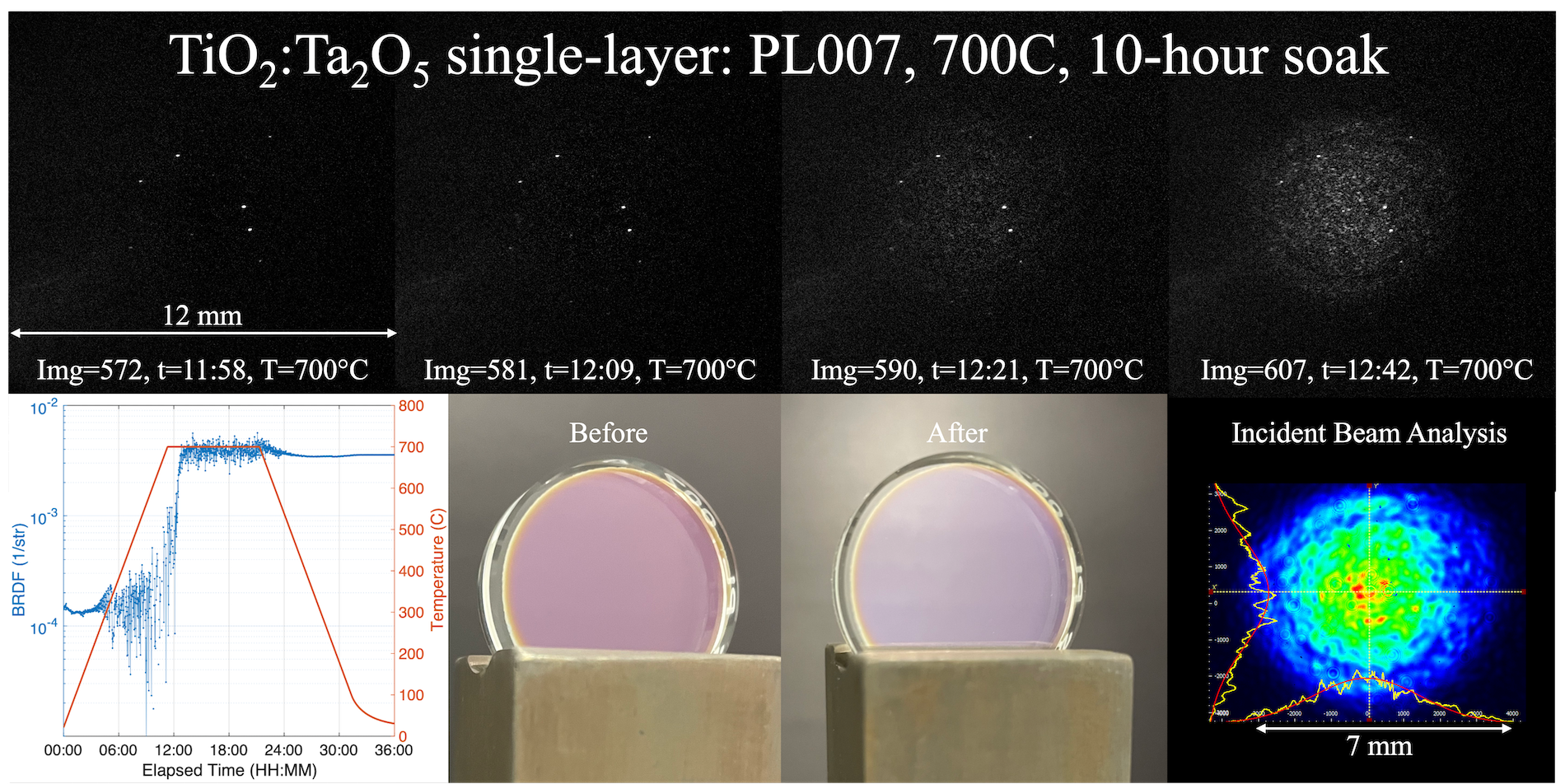}
  \caption{Results of annealing sample PL007, a single quarter-wavelength layer of TiO$_2$:Ta$_2$O$_5$ on a superpolished fused silica substrate. \textit{Bottom, left to right:} Measured oven temperature (orange) and BRDF (blue) of the sample, both versus elapsed time; sample before annealing; sample after annealing, showing the coating is slightly milky; intensity profile of the incident laser beam at the location of the sample, showing a beam diameter 7\,mm, and for this run, a more uniform Gaussian beam.
  The top row shows cropped images of a (12\,mm-wide) region of the sample illuminated by the SLD. The 700$^{\circ}$C soak starts at image 543 and elapsed time 11 hours, 21 minutes.
  \textit{Top, left to right:} Image 572, 37 minutes into the soak, shows several bright point scatterers are visible, but no sign of crystalization; Image 581, 48 minutes into the soak shows the first weak signs of crystallization; Image 590, one hour into the soak shows clear crystalization; Image 607, 81 minutes into the soak, the beam intensity profile is scattered quite uniformly by the crystallized coating. \label{fig:pl007-results}}
\end{figure}

Figure~\ref{fig:pl007-results} shows a similar composite image for sample PL007. For this sample, the two improvements mentioned above were both in place. The sample was held by a small stainless steel holder on a tall fused silica crucible, so the vertical translation due to thermal expansion was negligible. The iris used to pass the beam was enlarged so the beam incident on the sample was more uniform and Gaussian. The sample was ramped at 1$^{\circ}$C/minute and soaked at 700$^{\circ}$C for 10 hours. After the sample reaches its soak temperature at 11:20 elapsed time, a clear and gradually increasing crystallization is seen in the images and the BRDF lasting until 13:00 elapsed time, at which point no further increase in BRDF is seen. The visible light after image of the coating is again milky. 

\subsection{Side illumination example: TiO$_2$:GeO$_2$-coated sample 210811a} 

As described above, the AAS apparatus was found to have a second possible illumination and measurement channel in addition to the SLD front illumination. At elevated temperatures, thermal radiation from the heating elements provides a bright source of side illumination of the coatings. 
This light acts similarly to the light in a back- or side-illumination microscope and has proven particularly useful for viewing blisters and coating delamination, especially in the ``dark" images that are taken with the SLD off. 

\begin{figure}
  \centering
  \includegraphics[width=1.0\linewidth]{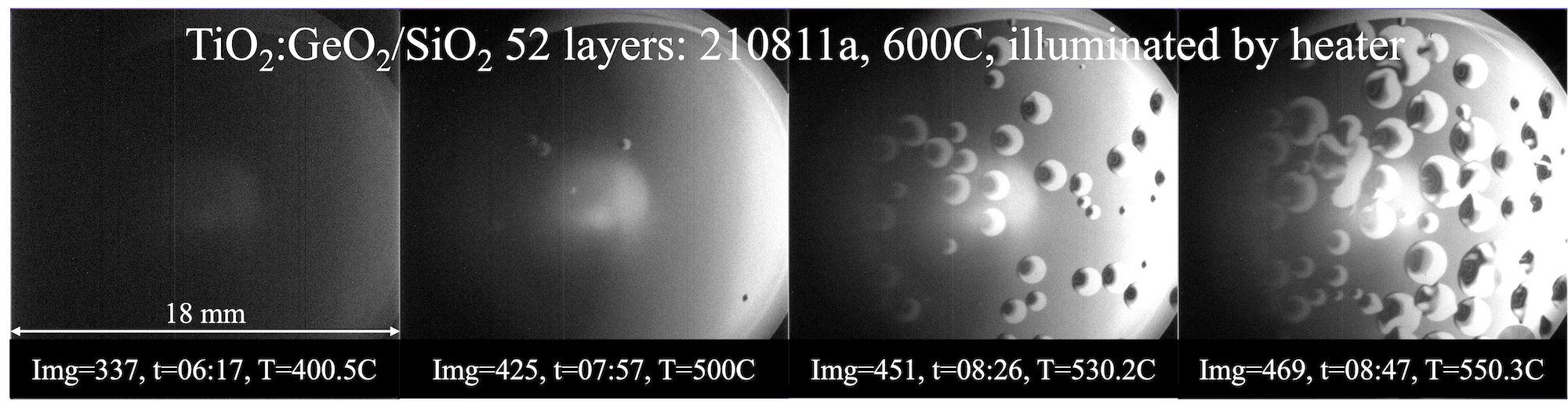}
  \caption{Results of annealing sample 210811a, a 52-layer quarter-wavelength stack of TiO$_2$:Ge$_2$ and Si$_{02}$ on a polished fused silica substrate. At 6 hours elapsed and 400$^{\circ}$C there are no blisters visible and the thermal radiation is visible but weak. At 500$^{\circ}$C small blisters have formed. At 530$^{\circ}$C the original blisters have grown and new blisters have formed and grown and begun to run into each other. At 550$^{\circ}$C the majority of the imaged surface is taken up by blisters.  \label{fig:210811a-results}}
\end{figure}

Figure~\ref{fig:210811a-results} shows a series of such dark images, for a test optic, from coating run 210811a, a 52-layer quarter-wavelength stack of TiO$_2$:Ge$_2$ and Si$_{02}$ coated by Carmen Menoni's group at Colorado State University on a polished fused silica substrate~\cite{Davenport:22}. 
The ramp rate is 1$^{\circ}$C/minute and the soak is 10-hours long. 
The images show the nucleation and growth of blisters, with one blister in the bottom right corner of the last image uncapping and ``popping off'' the coating.
Especially the onset of bubbling is much less visible, if at all, in the SLD-illuminated images. 
This coating performance is not indicative of the performance of titania-doped germania and silica coatings as this combination of rougher polish and many layers was known to lead to blisters. 
Such results provide insight into blister growth and delamination mechanisms, such as stress or outgassing, through the size and growth rate of blisters versus temperature. Such results will be published separately~\cite{Lalande}. 

\section{Conclusion}
\label{sec:conclusion}


We have described the Air Annealing Scatterometer and shown that it is capable of observing coating optical scatter, crystallization, and blisters/delamination throughout in-air annealing to 750$^{\circ}$C. 
The images and BRDFs presented here for TiO$_2$:Ta$_2$O$_5$ single-layer coatings with normal SLD illumination, reveal the onset, growth, and saturation (for given temperature profiles) of coating crystallization.
A potential follow-up project is to compare these optical results with x-ray diffraction to see which is most sensitive to the onset of crystallization. 
Because the data produced is in-situ, real time, and includes images and thus maps of the crystal formation over the coating, this method has advantages over other crystal measurement methods such as x-ray diffraction. 
Exemplary results that used an alternative light source, side illumination from the oven's heaters, were also presented, showing the formation and evolution of blisters and delamination in a TiO$_2$:Ge$_2$/Si$_{02}$ multilayer coating. 
These results will be explored more in future work and they suggest a possible upgrade to the AAS of dedicated side illumination. 

To our knowledge, the capabilities and type of results presented here have not been previously demonstrated in the literature.
Future steps will involve using the AAS to observe heat-induced damage mechanisms in candidate coating materials for future gravitational-wave detectors, especially those for which achievable annealing temperature appears to be limiting their performance. 
By providing onset and evolution data versus temperature, this apparatus will allow for deeper study of the damage mechanisms at play in coatings during high-temperature annealing. 
Such studies should lead to improvements in the coating manufacture process for low-optical-loss applications and thus improvements to future interferometric gravitational-wave detectors and atomic clocks.

Two main issues were identified with the setup in the course of this study. The first was thermal expansion of the stainless steel optic pedestal. This was solved by replacing the majority of the steel, except for a small optic holder, with a fused silica crucible. Results using this crucible showed negligible vertical translation and a BRDF trend without apparent translation artifacts. 
The second was a large variation in the BRDF caused by the heaters flashing via thermal radiation on minute timescales. This has not been solved yet. This could be addressed with radiation shields or further PID tuning.  


\begin{backmatter}
\bmsection{Funding}
Content in the funding section will be generated entirely from details submitted to Prism.

\bmsection{Acknowledgments}
Portions of this work were presented at the Optical Interference Coatings Conference in 2022, paper number ThB.3, entitled ``Imaging Scatterometer for Observing Changes to Optical Coatings During Air Annealing''~\cite{Rezac:22}. The authors thank the LIGO Scientific Collaboration Optics Working Group, especially Carmen Menoni (Colorado State), François Schiettekatte (Montreal), and Rana Adhikari (Caltech) for helpful discussions regarding this work. This work and the authors were supported by NSF grants PHY-2207998, PHY-1807069, AST-2219109, and AST-1559694, and by the Dan Black Family Trust and Nancy and Lee Begovich. AG was supported in part by Nancy Goodhue-McWilliams.  

\bmsection{Disclosures} The authors declare no conflicts of interest.

\bmsection{Data availability} Data underlying the results presented in this paper are not publicly available at this time but may be obtained from the authors upon reasonable request.


\end{backmatter}

\bibliography{AAS_bib}
\end{document}